\documentclass[twocolumn,amssymb,aps,prl,groupedaddress,showkeys,showpacs,10pt]{revtex4-1}
\usepackage{graphicx}
\usepackage{url}

\begin{document}

\title{Magnetogenesis Experiments Using A Modified Chaplygin Gas EoS}

\author{Patrick W. M. Adams}
\email[]{patrick.wm.adams@gmail.com}
\author{Bob Osano}
\email[]{bob.osano@uct.ac.za}

\affiliation{Astrophysics, Cosmology and Gravity Centre, Department of Mathematics and Applied Mathematics, University of Cape Town}

\date{\today}

\begin{abstract}
We examine magnetogenesis in a multi-fluid environment. We find that the various composition of a modified Chaplygin Gas (MCG) and Plasma Fluid (PF) yield magnetic fields of non-negligible strengths.These fields are produced by the battery effect and interactions between the two fluids may explain the amplification observed in the simulation. Our simulations show that the strongest fields are generated in a mixture with 50$\%$ MCG and 50$\%$ PF.					                      
\end{abstract}



\maketitle

\section{Introduction}
          
Magnetic fields pervade the observable part of the universe. They exist on both the astrophysical and cosmological scales. The strengths of these fields vary with scale, for example fields with microgauss strength and coherence scales of order kiloparsecs are observed in our neighbourhood of galaxies and in recently-formed galaxies, while fields of even higher microgauss, having greater coherence scales, are observed in clusters of galaxies \cite{brandenburg2005}. The origin of the fields is poorly understood at present, and remains an important and unresolved problem in astrophysics and cosmology. Several generative mechanisms have been proposed (for a general review on cosmic magnetogenesis, see \cite{hogan1983, quashnock1989, vachaspati1991}). These mechanisms are divided into two categories that are distinguishable by when they operate. Mechanisms of an astrophysical nature operate at the large-scale structure formation, or later, and may be the result of the Biermann battery effect \cite{biermann2003}, the first supernova remnants \cite{miranda1998, hanayama2005, hanayama2009}, the Weibel instability \cite{weibel1959}, or even of the intergalactic plasmas \cite{lazar2009}. On the other hand, mechanisms may operate in the early universe which may also be subdivided into those operating during an inflationary epoch of the universe \cite{turner1988, ratra1992, bassett2001, campanelli2009, byrnes2012,fer2013,fer2014}, and those operating after inflation has ended \cite{harrison1970, dolgov2001, berezihani2004, betschart2004, matarrese2005, jimenez2011}, such as cosmological phase transitions \cite{hogan1983, sigl1997, stevens2012}. The gas in galaxies, as in other astrophysical environments, is either fully- or partially-ionized. This allows it to posses electric currents, which according to Maxwell's equations, produces magnetic fields. Physically, the accompanying Lorentz force acts on the plasma (ionized gas) thereby producing the effect via the momentum equation for the plasma. These interactions of magnetic fields and plasma (treated as a fluid) are studied in Magnetohydrodynamics (MHD). In MHD, Maxwell's equations of electrodynamics are combined with fluid equations, where the Lorentz forces due to electromagnetic fields are included.

\section{The MHD Equations}
Studies of MHD involve investigations of how a fluid that is made up of charged particles moves in the presence of electromagnetic fields. Several approximations and assumptions are made: (i) the properties of the particles are averaged over volumes that are much smaller than macroscopic volumes but larger than inter-particle distance, (ii) there is charge neutrality to zeroth order, (iii) the characteristic frequencies are much less than the lowest ion gyro-frequencies, (iv) the difference in mean velocities of individual species of particles is small compared to the fluid velocity. With these assumptions it can be shown that the set of equations that govern dynamics of such fluids is:
\begin{eqnarray}
\frac{1}{\rho}\frac{\mathrm{D}\rho}{\mathrm{D}t} & = & -\nabla\cdot\mathbf{v} \label{eq:Icont} \\
\frac{\mathrm{D}\mathbf{v}}{\mathrm{D}t} & = & -\frac{\nabla p}{\rho} - \nabla\Phi_{\mathrm{grav}} + \frac{\mathbf{J}\times\mathbf{B}}{\rho\mu_0} + \nu\mathbf{f}_\mathrm{v} + \frac{\mathbf{f}_\mathrm{b}}{\rho} \label{eq:Ins} \\
\frac{\partial\mathbf{B}}{\partial t} & = & \nabla\times\left(\mathbf{v}\times\mathbf{B} - \eta\mu_0\mathbf{J}\right) \label{eq:Iind} \\
\frac{\mathrm{D}e}{\mathrm{D}t} & = & \frac{\eta}{\rho}\mathbf{J}^2 - (\gamma-1)e\nabla\cdot\mathbf{v}, \label{eq:Ient}
\end{eqnarray}
where $\mathbf{\rho}$ is the fluid density, $\mathbf{v}$ is the fluid velocity, $\mathbf{B}$ is magnetic flux density, and $e$ is the internal energy.  $\mathrm{D}/\mathrm{D}t$ is the Lagrangian derivative, $\mathbf{J}$ is the current density, $\gamma$ is the ratio of specific heats, $\eta$ is the magnetic resistivity, $\mu_0$ is the vacuum permiability, $\Phi_{\mathrm{grav}}$ is gravitational potential, the term $\mathbf{f}_\mathrm{v} = \left(\nabla^2\mathbf{v} + \frac{1}{3}\nabla\nabla\cdot\mathbf{v}\right)$ are the viscous forces, while the term $\mathbf{f}_\mathrm{b}$ subsumes all additional unaccounted-for body forces acting on the fluid. In order to complete the set, we need to provide an equation of state for the fluid, which could be in the barotropic form, $p = p(\rho)$, and will also demand that the divergenceless condition, $\nabla\cdot\mathbf{B} = 0$, be satisfied. Studies of the generation and evolution of magnetic fields often involves variations of the induction equation given by Eq. (\ref{eq:Iind}). In the foregoing approximation, Amp\`ere's Law, $\mu_{0}\mathbf{J}=\nabla\times\mathbf{B}$, can used to eliminate the current density, leading to:
\begin{eqnarray}
\frac{\partial\mathbf{B}}{\partial t} & = & \nabla\times\left(\mathbf{v}\times\mathbf{B} - \eta\nabla\times\mathbf{B}   \right) \label{eq:Iind2},
\end{eqnarray}
given the standard Ohm's law. The first term on the RHS (the induction term) could be expanded, as required, using vector identities. The important point to take note of is that  $\mathbf{B}=0$ is a solution to this Induction Equation. In order to generate magnetic fields from zero, one needs to find away of violating this equation. The simplest and most natural way of doing this is to add a battery term. In our study, we consider a multi-fluid environment that includes plasma and the Chaplygin Gas. The Chaplygin Gas interacts gravitational with plasma, which can lead to a situation where the electron density gradient is not parallel to the temperature gradient, as found in the thermally generated seed fields \cite{sub177} in the cosmological context, or even in cosmological shocks \cite{kuls178,davies186}. We first discuss the Chaplygin Gas.

\section{The Chaplygin Gas}
About 70$\%$ of the total energy of the universe is in the form of dark energy  as observed in the CMBR \cite{CMB}, SDSS \cite{dbSDSS,hanaSDSS,bacSDSS,perlSDSS} and SNIa \cite{SNIa} experiments. This energy is thought to be responsible for the present acceleration in the expansion of the universe. Among the candidates that have been proposed for dark energy, and which have been confronted with observation, is the generalized Chaplygin Gas \cite{bento2002, bento2003}. It is conjectured that dark energy and dark matter could be unified by using the CG's exotic equation of state (EoS). Its utility lies in the fact that it is able to interpolate naturally between the dark energy- and dark-matter-dominated eras of the universe (e.g. \cite{tupper2009})

The Chaplygin Gas is a hypothetical substance that was proposed by Sergei Chaplygin in 1902 in his work on Gas Jets \cite{chaplygin}. It possesses several interesting features, one namely being that it is able to behave as pressureless fluid when the value of the scale factor is small, but like a cosmological constant when the value of the scale factor is large, which leads to the acceleration of the expansion of the universe \cite{ben2002}.

The Chaplygin gas EoS relates the pressure, $p$, to the density, $\rho$, in the exotic form:

\begin{equation}	                  
\label{eq:chapeos}
p = -\frac{1}{\rho}.
\end{equation}

We shall consider a modified version of this EoS, given by:

\begin{equation}	                  
\label{eq:chapeos2}
p = \mathcal{A}\rho -\frac{\mathcal{B}}{\rho^\alpha},
\end{equation}

where $\mathcal{A}$ and $\mathcal{B}$ are positive constants, and $0\leq\alpha\leq1$. For our purposes, and in line with WMAP and SDSS observations, we choose the value of $\mathcal{A} = 0.085$, and $\alpha = 1.724$ \cite{lu2008,ranjit2014}, whilst $\mathcal{B}$ shall be assumed to be unity. 

\section{The Plasma Fluid}
The plasma fluid is taken to be that of ideal plasma. We consider this two-fluid component, with the electrons and ions as separate fluids that interact through collisions. For a simple picture, we take ions as having one charge (i.e. they are just protons). We take the stress tensor to be just the isotropic pressure, leaving out any non-ideal terms, but adopt a simple collision between the particle species. It follows that the electrons and ions will have separate equations of motion that take the form of Eq. (\ref{eq:Ins}). It can be shown, in the limit in which the ratio $\mathbf{m}_e/\mathbf{m}_i$ is much less than 1, that the differences in the two equations of motions yield:
\begin{eqnarray}
(\mathbf{E}+\mathbf{v}_\mathrm{i}\times\mathbf{B})=-\frac{\nabla p_\mathrm{e}}{e n_\mathrm{e}}+ \Gamma,\label{eq:ohm}
\end{eqnarray}
where subscript $\mathrm{i}$ and $\mathrm{e}$ stand for ions and electrons respectively. $\Gamma$ represents the sum of the Ohmic terms, Hall's electric fields and inertial terms. The first term on the RHS of Eq. (\ref{eq:ohm}) is the Biermann battery term, which leads to thermally-generated electromagnetic fields. We are interested in the case where this term has a curl, for only then will it be able to produce a magnetic field. We have so far discussed the fluid behaviour for the positive and negative particles, in the absence of other components. Note that the neutral particle will play a role if the plasma is only partially ionized. More interesting is that the presence of Chaplygin Gas will change the dynamics of the two fluids and the way they interact, as the two species respond differently to the gravitational effect of the gas. This in turn will affect Eq. (\ref{eq:ohm}) and, in particular, the battery term. It is this effect and how this affects the growth of magnetic field that we simulate in this article.

We treat both the fluid pressure and density as nett quantities, i.e.:

\begin{eqnarray}
\label{eq:pddecomp}
p & = & p_{\mathrm{PF}} + p_{\mathrm{MCG}} \nonumber \\
\rho & = & \rho_{\mathrm{PF}} + \rho_{\mathrm{MCG}}, \nonumber
\end{eqnarray}
where PF is the Plasma Fluid component, and MCG is the modified Chaplygin Gas component, leading to an EoS that reflects the presence of both substances. In the context of cosmology, $p_\mathrm{MCG}$ would represent the pressure contribution of dark energy only, as the pressure contribution of dark matter is zero. The form of the EoS may then be given by:

\begin{equation}
\label{eq:simeos}
p = \mathcal{A}\rho - \frac{\mathcal{B}}{\rho^\alpha} + \mathcal{C}\rho,
\end{equation}
which is the form that we use for our simulations. Furthermore, it may also be assumed that both the PF and MCG together comprise 100\% of the fluid under consideration. Thus, their relative constituency within the fluid may be expressed via a simple percentage relationship (e.g. 50\% (PF)/50\% (MCG), 80\% (PF)/20\% (MCG), and so forth).

\section{Simulation Results and Discussion}

All simulations were done on a $32^3$ periodic box of length $L = 2\pi$, using the \textsc{Pencil Code}\footnote{The code may be obtained at the following URL: \protect\url{http://pencil-code.nordita.org/}}, which is a high-order finite difference code used for performing numerical simulations of charged fluids, as well as other MHD phenomena \cite{pencilman}.

Simulations conducted considered three cases of a fluid consisting of both PF and MCG components in the percentage makeup of: 90\% (PF)/10\% (MCG), 50\% (PF)/50\% (MCG), and 10\% (PF)/90\% (MCG). These results were compared to the case with a fluid consisting solely of the PF, as well as a fluid consiting solely of the MCG with $\mathcal{A} = 0.085$, $\mathcal{B} = 1$ and $\alpha = 1.724$. In these simlations, the magnetic field was allowed to grow from zero initial conditions, which was facilitated by the presence of a numerical forcing term present in the Navier-Stokes equations; the magnetic ionization factor, $\chi$, was also set to zero, representing a fully-ionized plasma.

As the \textsc{Pencil Code} simulates quantities that are unit-agnostic, the results may be interpreted appropriately depending on the physical context. In particular, should they be interpreted within a cosmological context, one would have to take into account that the magnetic flux strength, $\mathbf{B}$, could in fact refer to a rescaled version of the actual magnetic flux strength (e.g. $\mathbf{\widetilde{B}} = \mathbf{B}/a^2$, where $a$ is the scale factor from the Friedmann Equation) that also takes into account the effects of the expanding universe. In order to obtain the evolution of the true magnetic flux strength, then, the value of the scale factor at the era in which the result is to be interpreted in must be known before any other physics can be discussed. 

\begin{figure*}[htbp]
	\centering
		\includegraphics[scale=0.8,viewport =50 240 550 620,clip]{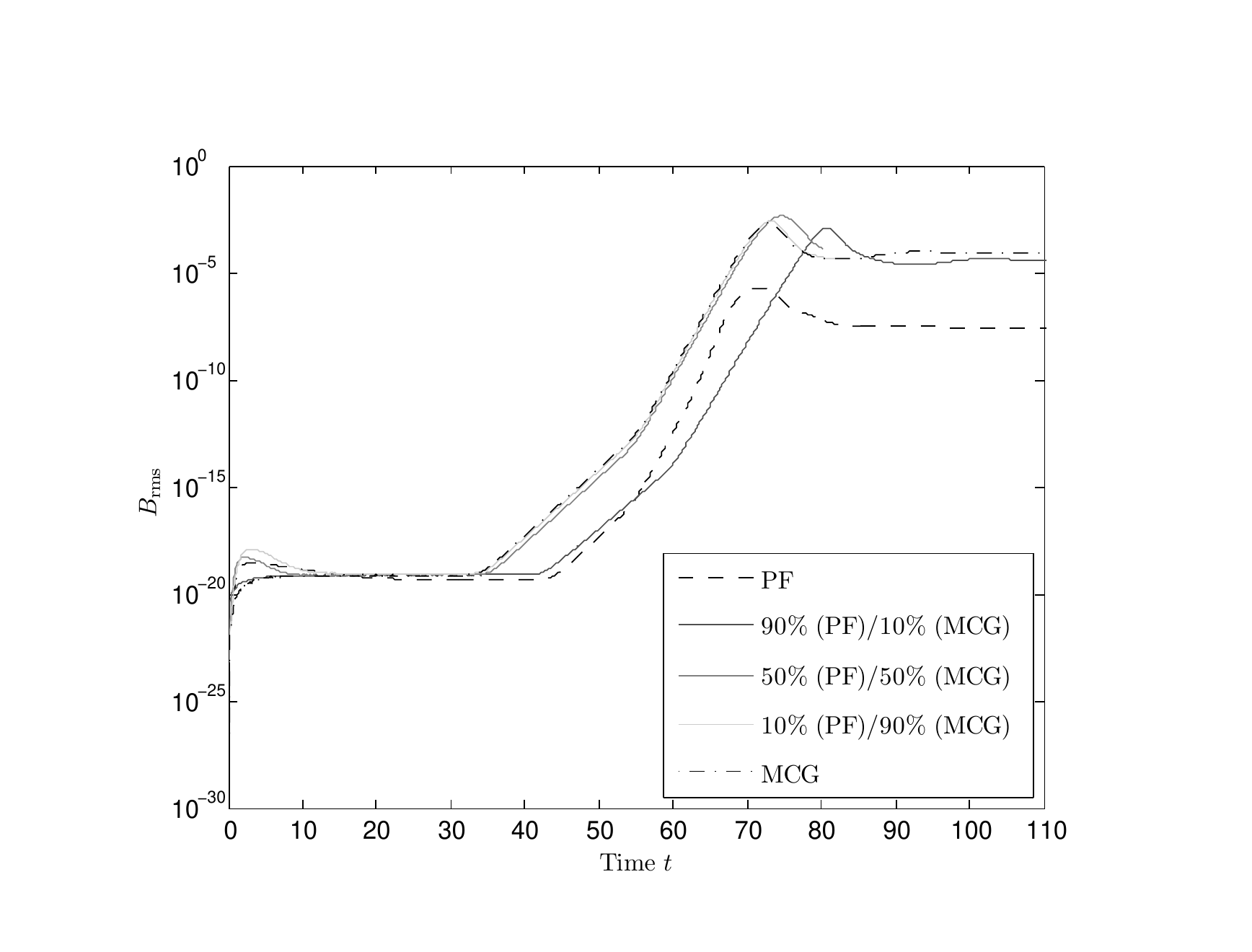} 
	\caption{Full-time evolution of $B_\mathrm{rms}$ for each of the plasmas simulated. A rapid exponential growth of the field strength to a global maximum is evident in all cases. At later times, the field strength drops slightly and settles down to a relatively constant value. Note that the PF-only plasma produces the weakest fields, whilst the 50\% (PF)/50\% (MCG)-mix produces the strongest fields.}
	\label{fig:plotfull}
\end{figure*}

In order to quantify the growth of the magnetic fields, we chose to follow the temporal evolution of the rms strength of the magnetic flux density, $B_{\mathrm{rms}}$ (hereafter, simply referred to as the magnetic field strength); these results are displayed in figure \ref{fig:plotfull}.

\begin{figure*}[htbp]
	\centering
		\includegraphics[scale=0.80,viewport =50 240 550 620,clip]{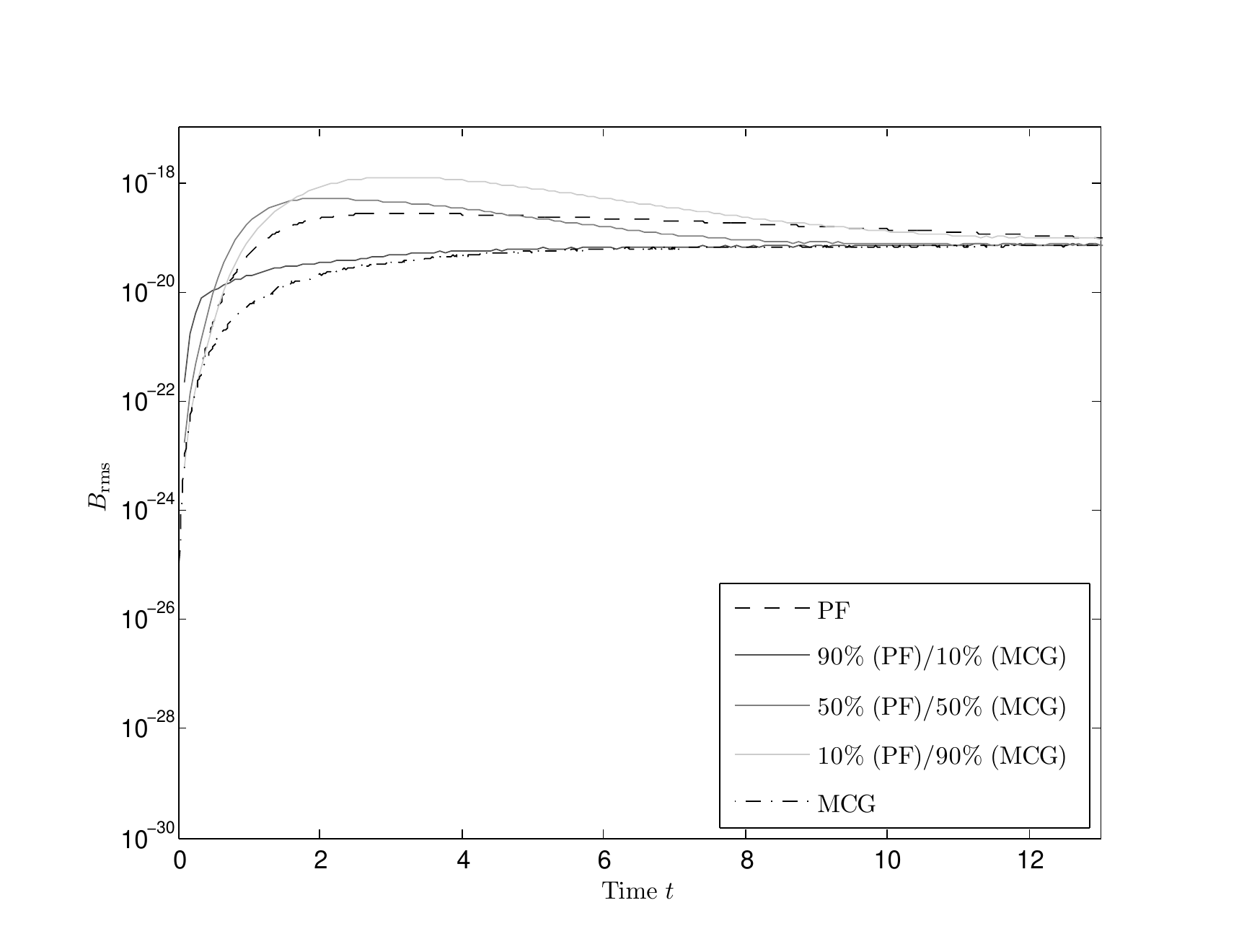} 
	\caption{Early-time evolution of $B_\mathrm{rms}$ for each of the plasmas simulated. Here a rapid initial upshoot and then gradual exponential decrease in strength is observed in all cases.}
	\label{fig:plotearly}
\end{figure*}

Examining the growth of the magnetic field strength at early times (see figure \ref{fig:plotearly}), it is evident that the field immediately enters an exponential growth phase and grows up to a local maximum value, after which it then gradually drops off exponentially in strength once more. After a finite time, the field then enters a second major exponential growth phase, this time increasing in strength by many orders of magnitude, reaching a global maximum value, after which it again drops exponentially in strength, eventually settling on a roughly constant value. Long runs conducted specifically for observing of late-time behaviour suggest that this second settled phase of the field strength is permanent, though confirmation of this may only be obtained by examining the magnetic energy spectra in detail. 

As shown in figure \ref{fig:plotfull}, we may conclude that a mixture consisting of 50\% (PF)/50\% (MCG) produces the strongest magnetic fields, with the other MCG-containing mixtures being close contenders; an PF-only plasma apparently produces the weakest magnetic fields.

\begin{table}
	\centering
		\begin{tabular}{ccc}
			\hline
			\hline
			Mixture & $t_\mathrm{max}$ & $B_\mathrm{rms}$ \\ \hline
			PF & 71.9627 & $1.914744\times10^{-6}$ \\
			90\% (PF)/10\% (MCG) & 80.7623 & 0.001219757 \\
			50\% (PF)/50\% (MCG) & 74.6370 & 0.004417391 \\
			10\% (PF)/90\% (MCG) & 73.3676 & 0.002724538 \\
			MCG & 73.1455 & 0.002128081 \\
			\hline
			\hline
		\end{tabular}
	\caption{Global maximum values of $B_\mathrm{rms}$ and the approximate times at which they occured.}
	\label{tab:bstrength}
\end{table}

In table \ref{tab:bstrength}, we present the present the global maxima of the fields generated by each mixture, along with the approximate times that these occured at. It is once more clear that the above-mentioned  50\%/50\% mixture produces the strongest magnetic fields. Examining this table together with figure \ref{fig:plotfull}, however, reveals a rather interesting phenomenon: though all of the mixtures reached their global maxima around roughly the same time (around $t \sim 71 - 73$), only the 90\% PF mixture appears as a stark outlier, reaching its global maximum around $t\sim80$!

It is also clear that each of these fields experience major growth levels during the simulations; examining the data for the 10\% (PF)/90\% (MCG) mixture suggests that the field strength grows by $\sim10^6\%$ during the primary exponential growth phase, and then grows by another $\sim10^{18}\%$ when the secondary exponential growth phase is complete. Similar growth levels were observed for field strengths arising from the other mixtures.

For the remaining discussions, we shall only consider the behaviour of the magnetic fields produced from the 50\% (PF)/50\% (MCG) mixture. As can be seen in figure \ref{fig:plotfull}, and may be inferred from the following figures, the behaviour of the fields arising from the other mixtures is qualitatively identical.

Along with the temporal evolution of the magnetic field strength, we also tracked the evolution of the spectral magnetic energy across the 32 wavenumbers present within the system, which correspond to the number of mesh points used. This was done in order to assess the flow of spectral energy across the large and small scales, which are represented by smaller and larger values of the wavenumbers, $k$, respectively.

\begin{figure*} 
	\centering
		\includegraphics[scale=0.80,viewport =50 240 550 620,clip]{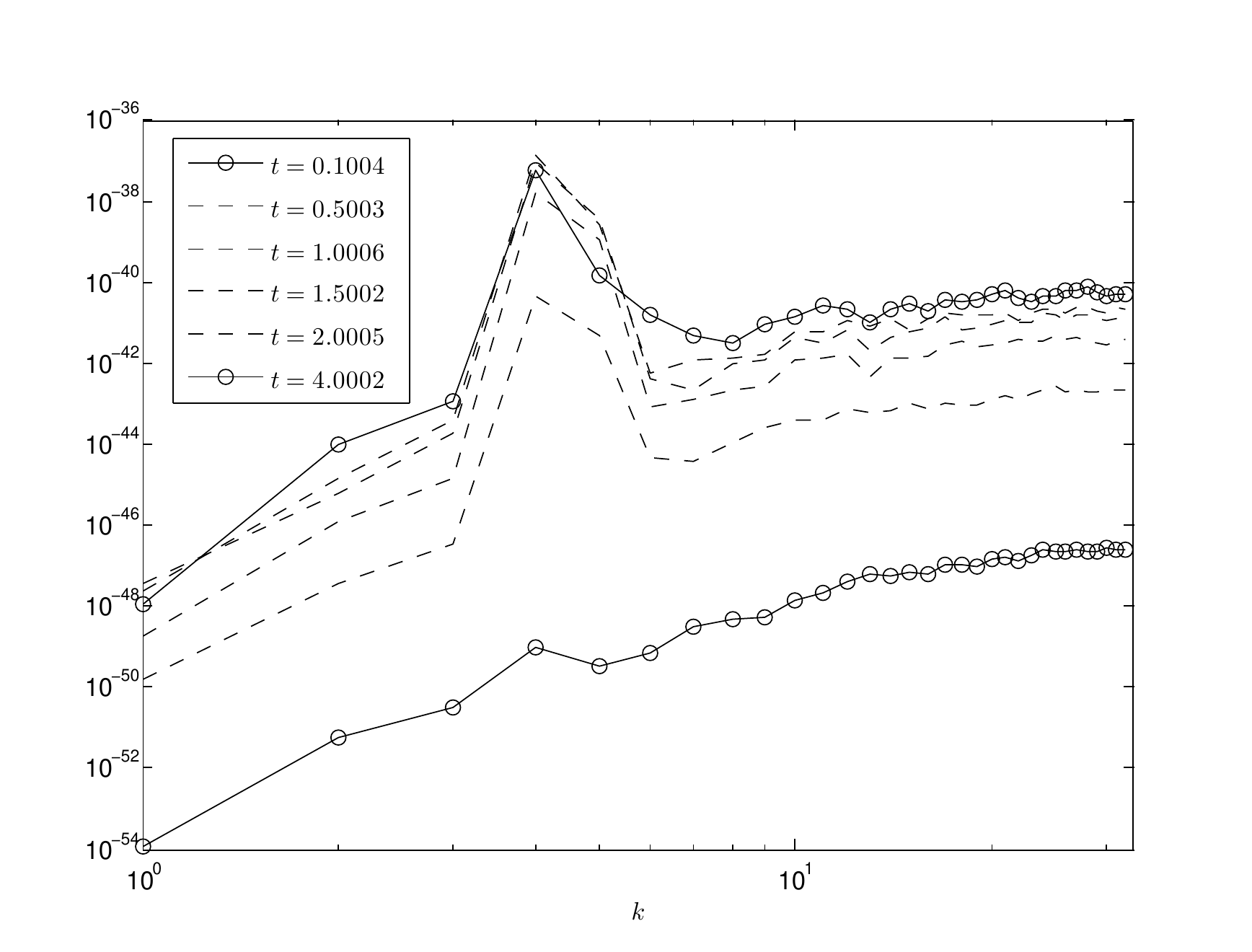} 
	\caption{Early-time growth of the magnetic energy spectra for the 50\% (PF)/50\% (MCG) plasma. Note the sharp peak at the wavenumber $k=4$. All other plasmas simulated display identical qualitative behaviour.}
	\label{fig:5050espec}
\end{figure*}

\begin{figure*} 
	\centering
		\includegraphics[scale=0.80,viewport =50 240 550 620,clip]{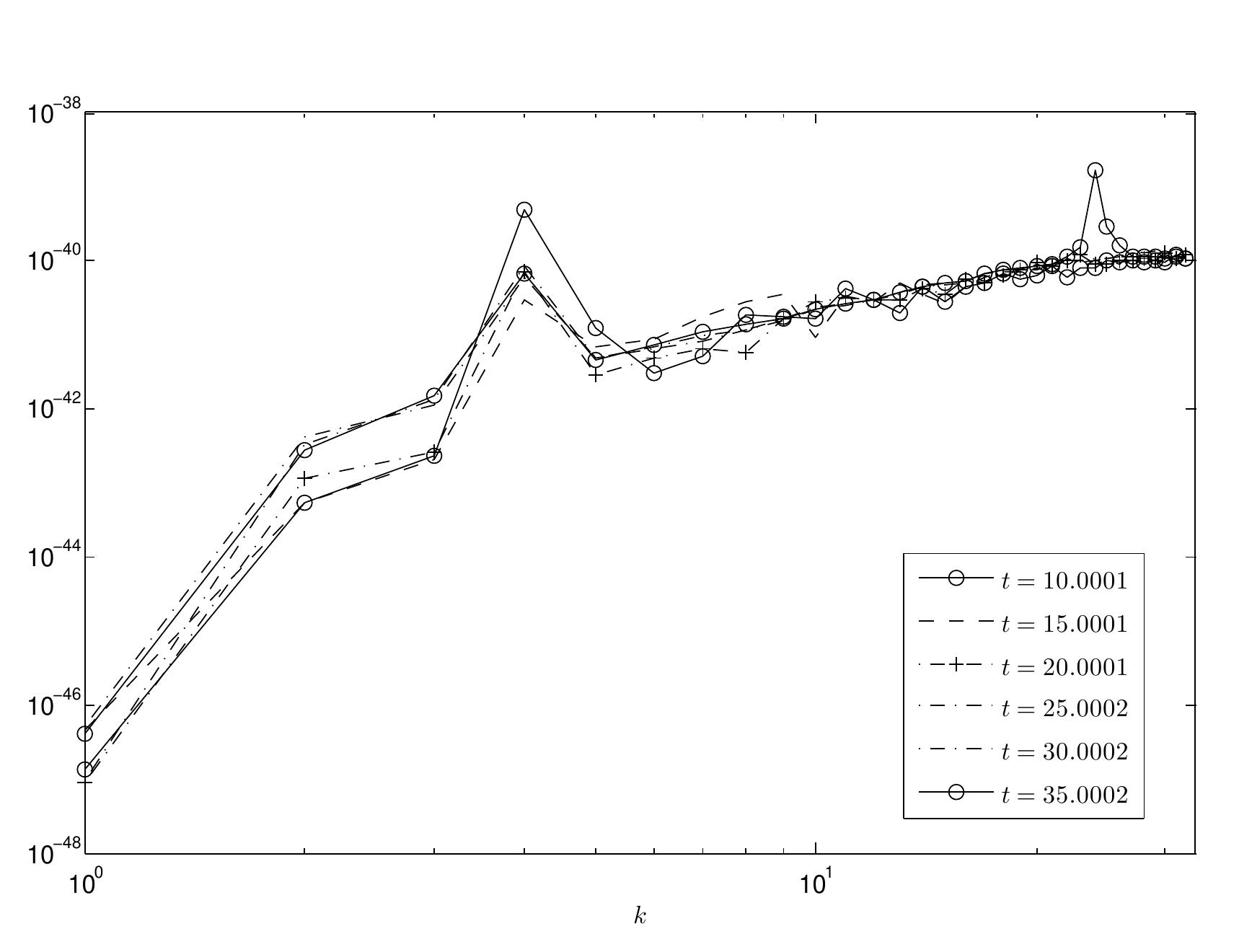} 
	\caption{Growth of the magnetic spectral energy during the initial exponential decay for the 50\% (PF)/50\% (MCG) plasma. Note how the peak at $k = 4$ subsides gradually over time as energy moves to the higher wavenumbers. Before the exponential growth phase begins, a sharp peak is now observed at $k = 24$.}
	\label{fig:5050mspec}
\end{figure*}

\begin{figure*} 
	\centering
		\includegraphics[scale=0.80,viewport =50 240 550 620,clip]{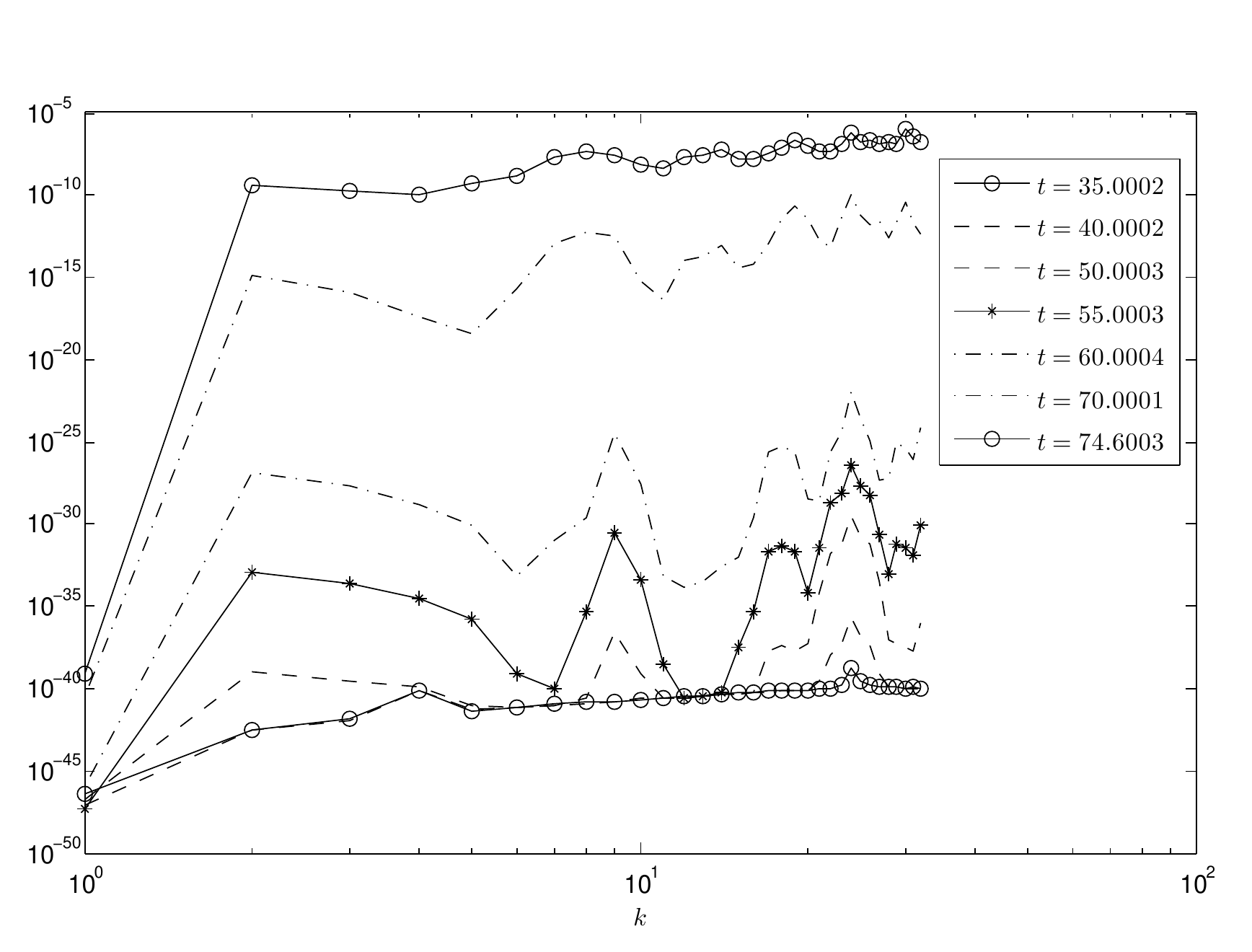} 
	\caption{Growth of the magnetic spectral energy during the major exponential growth phase. Note how the peak at $k = 24$ consistently subsides over time as energy moves back to the lower wavenumbers, creating additional shrinking peaks as the spectrum grows.}
	\label{fig:5050lspec}
\end{figure*}

Examining figure \ref{fig:5050espec}, which corresponds to the early inital growth phase of the field strength shown in figure \ref{fig:plotearly}, we note a sharp increase in magnetic spectral energy at wavenumber $k = 4$, with energy appearing to move rapidly down to the smaller scales. During the initial decline of the field strength (figure \ref{fig:5050mspec}), after reaching the first maximum value, the peak observed at $k = 4$ appears to subside, whilst the energy at the other scales still appears to be growing very slowly. This behaviour continues until the begnning of the second major exponential growth phase of the field, which is marked clearly by a sharp peak at wavenumber $k = 24$. The entirety of the second growth phase is then dominated by the sharp peak at $k = 24$, which appears to shrink rapidy as time passes and energy is distributed back to the larger scales, also creating several other peaks in the process. Eventually, the spectrum appears to begin settling at all scales, except at $k=1$ which is still growing slowly, corresponding to the late-time settling of the field strength observed in figure \ref{fig:plotfull}.

From this, it is evident that magnetic spectral energy is being transferred across the large and small scales by some mechanism. Consistent and rapid growth of the spectra during the magnetic field's exponential growth phases suggest that these phases themselves could possibly be attributed to the operation of a kinematic dynamo which eventually becomes saturated after the field reaches its maximum strength.

\section{Conclusion}

This paper investigated magnetogenesis in a two-fluid toy model involving the Ideal and Chaplygin gases. Simulations performed looked at the growth of the magnetic field arising from a plasma consisting of varying percentage contributions of both of the aforementioned gases; in particular, mixtures consisting of 10\% (PF)/90\% (MCG), 50\% (PF)/50\% (MCG) and 90\% (PF)/10\% (MCG) were used. These were also compared to magnetic fields grown in simulations run on plasmas consisting only of the Plasma Fluid and Chaplygin Gas respectively. It was found that the 50\% (PF)/50\% (MCG) mixture gave rise to the strongest magnetic fields, whilst the case of a 100\% PF plasma gave rise to the weakest fields. All of the fields simulated displayed qualitatively identical features during their evolution, namely exponential growth followed by exponential decay to a local minimum at very early times, and then immediate exponential growth to a global maximum once this local minimum was reached. Beyond this, slight exponential decay was once more observed, as the field eventually settled down to an apparently constant strength for the remainder of the run.  An investigation of the magnetic energy spectra suggested that magnetic spectral energy was being transferred across the large and small scales as the field grew in strength, suggesting that a kinematic dynamo may be operating in the phases of the field's exponential growth; this dynamo was then seen to saturate, which could explain the second minor exponential decay and settling of the field strength.

We note that before any further physics can be discussed, the exact contributions of each of the terms in the Induction and (especially) the Navier-Stokes equations to the growth of the magnetic field strength must be quantified. This calls for further simulations to be conducted so that the model may be refined.

\begin{acknowledgments}

\centerline{$\bigstar$}
The authors wish to thank Prof. Axel Brandenburg of the Nordic Institute for Theoretical Physics (NORDITA) for technical help with the \textsc{Pencil Code}.

The authors also wish to acknowledge the financial support of the National Research Foundation of South Africa (NRF), administered by the Postgraduate Funding Office of the University of Cape Town, that made this research possible.
\end{acknowledgments}

\appendix*
\section{Appendix: The \textsc{Pencil Code}}\label{app:pencil}

The \textsc{Pencil Code} is a high-order finite difference code used for performing numerical simlations of charged fluids and other magnetohydrodynamic phenomena. It solves the standard equations of compressible MHD for the density as stated above, now given by:

\begin{widetext}
\begin{eqnarray}
\frac{\mathrm{D}\ln\rho}{\mathrm{D}t} & = & -\nabla\cdot\mathbf{v} \label{eq:cont} \\
\frac{\mathrm{D}\mathbf{v}}{\mathrm{D}t} & = & -c_s^2\gamma\nabla\left(\frac{s}{c_p} + \ln\rho\right) - \nabla\Phi_{\mathrm{grav}} + \frac{\mathbf{J}\times\mathbf{B}}{\rho} + \nu\left[\nabla^2\mathbf{v} + \frac{1}{3}\nabla\left(\nabla\cdot\mathbf{v}\right) + 2\mathbb{S}\cdot\nabla\ln\rho\right] + \zeta\nabla(\nabla\cdot\mathbf{v}) \label{eq:ns} \\
\frac{\partial\mathbf{A}}{\partial t} & = & \mathbf{v}\times\mathbf{B} - \eta\mu_0\mathbf{J} + \frac{1}{1+\chi}\frac{\nabla p}{\rho} \label{eq:ind} \\
\varrho T\frac{\mathrm{D}s}{\mathrm{D}t} & = & \mathcal{H} - \mathcal{C} + \nabla\cdot\left(K\nabla T\right) + \nu\mu_0\mathbf{J}^2 + 2\rho\nu\mathbb{S}\otimes\mathbb{S} + \zeta\rho(\nabla\cdot\mathbf{v})^2. \label{eq:ent}
\end{eqnarray}
\end{widetext}

Note once more that all fields and scalar functions are functions of space and time, $(\mathbf{x},t)$. Here, $\Phi_\mathrm{grav}$, a gravitational potential, $\nu$, the kinematic viscosity, $\zeta$, a so-called shock viscosity, $\eta$, the magnetic diffusion, $\mu_0$, the magnetic permeability, $\chi$, the degree of ionization, $T$, the temperature, $\mathcal{H}$ and $\mathcal{C}$ explicit heating and cooling terms, and $K$, the radiative thermal conductivity. $\mathbb{S}$ is the traceless rate-of-strain tensor, given as:

\begin{equation}
\label{eq:straintens}
\mathbb{S}_{ij} = \frac{1}{2}\left(\frac{\partial u_i}{\partial x_j} + \frac{\partial u_j}{\partial x_i} - \frac{2}{3}\delta_{ij}\nabla\cdot\mathbf{v}\right) \nonumber
\end{equation}

in Cartesian co-ordinates, and $c_s^2$ is the squared sound-speed, given by:

\begin{equation}
\label{eq:csound}
c_s^2 = \gamma\frac{p}{\rho} = c_{s_0}^2\exp\left[\frac{s\gamma}{c_p} + (\gamma - 1)\ln\frac{\rho}{\rho_0}\right], \nonumber
\end{equation}

where $c_{s_0}$ is a reference value of the speed of sound, taken at some reference height; all other symbols retain their prior meanings. We have also made the identification that $\nabla\times\mathbf{A} = \mathbf{B}$, the magnetic flux density\footnote{This automatically implements the no-divergence condition on $\mathbf{B}$ as required by Gauss' Law for Magnetism.}, and that $\nabla\times\mathbf{B}/\mu_0 = \mathbf{J}$, the current density. 

As we are studying the generation of magnetic fields from zero initial conditions, we must also couple to the Induction Equations a battery term in order to allow any fields to grow. For the purposes of this work, we use the Biermann Battery to generate the magnetic fields, inserting the relevant term into the Induction Equations above.

\bibliography{chapmsc}

\end{document}